\documentclass[a4paper]{article}
\usepackage{amsmath,graphicx}
\usepackage{INTERSPEECH2022}
\usepackage{url}
\usepackage{booktabs}
\usepackage{hyperref}
\usepackage{xcolor}
\usepackage{multicol}

\usepackage{bm}

\newcommand{\hB}{\mathbf{h}}

\newcommand{\xB}{\mathbf{x}}

\newcommand{\zB}{\mathbf{z}}

\newcommand{\LM}{\mathcal{L}}

\newcommand{\Sec}[1]{section~\ref{sec:#1}}
\newcommand{\Fig}[1]{Fig.~\ref{fig:#1}}
\newcommand{\Table}[1]{Table~\ref{tbl:#1}}

\newcommand{\method}[1]{\textbf{#1}}
\newcommand{\mysection}[1]{\vspace{-5pt}\section{#1}\vspace{-3pt}}
\newcommand{\mysubsection}[1]{\vspace{-4pt}\subsection{#1}\vspace{-2pt}}
\newcommand{\mysubsubsection}[1]{\vspace{-2pt}\subsubsection{#1}\vspace{-2pt}}
\newcommand{\revise}[1]{\textcolor{black}{#1}}
\newcommand{\cameraready}[1]{\textcolor{black}{#1}}

\title{MSR-NV: Neural Vocoder Using Multiple Sampling Rates}

\name{Kentaro Mitsui, Kei Sawada}
\address{rinna Co., Ltd., Japan}
\email{kemits@rinna.co.jp, keisawada@rinna.co.jp}

\begin{document}

\setlength{\abovedisplayskip}{2pt}
\setlength{\belowdisplayskip}{2pt}
\maketitle

\begin{abstract}
The development of neural vocoders (NVs) has resulted in the high-quality and fast generation of waveforms.
However, conventional NVs target a single sampling rate and require re-training when applied to different sampling rates.
A suitable sampling rate varies from application to application due to the trade-off between speech quality and generation speed.
In this study, we propose a method to handle multiple sampling rates in a single NV, called the MSR-NV.
By generating waveforms step-by-step starting from a low sampling rate, MSR-NV can efficiently learn the characteristics of each frequency band and synthesize high-quality speech at multiple sampling rates.
It can be regarded as an extension of the previously proposed NVs, and in this study, we extend the structure of Parallel WaveGAN (PWG).
Experimental evaluation results demonstrate that the proposed method achieves remarkably higher subjective quality than the original PWG trained separately at 16, 24, and 48 kHz, without increasing the inference time.
We also show that MSR-NV can leverage speech with lower sampling rates to further improve the quality of the synthetic speech.
\end{abstract}


\section{Introduction}
\label{sec:intro}
In text-to-speech synthesis, singing voice synthesis, and music synthesis, studies to improve the quality of vocoders, which generate waveforms from acoustic features, have been extensively conducted.
Neural vocoders (NVs), which use neural networks to generate waveforms, have greatly improved the quality of synthetic audio.
WaveNet~\cite{oord2016wavenet, tamamori2017speaker}, a representative of autoregressive (AR) NVs, achieved a significantly higher quality than conventional signal processing-based vocoders~\cite{kawahara2008tandem, morise2016world} by predicting waveform samples individually.
Although AR NVs can generate high-quality audio because they can use previous prediction results, they suffer from a slow generation speed.
To address this issue, fast and high-quality waveform generation using non-autoregressive (non-AR) NVs has been actively studied.
Various generative models have been used in this area, including 
inverse autoregressive flow (IAF)~\cite{kingma2016iaf} used in Parallel WaveNet~\cite{oord2018parallel} \revise{and ClariNet~\cite{ping2019clarinet}},
generative flow (Glow)~\cite{kingma2018glow} used in WaveGlow~\cite{prenger2019waveglow}, the
generative adversarial network (GAN)~\cite{goodfellow2014gan} used in several NVs~\cite{kumar2019melgan, yamamoto2020parallel, yang2020vocgan, kong2020hifigan},
and the denoising diffusion probabilistic model (DDPM)~\cite{ho2020ddpm} used in WaveGrad~\cite{chen2021wavegrad} and DiffWave~\cite{kong2021diffwave}.
\revise{Moreover, several methods incorporating signal processing insights have been proposed, including the neural source-filter model~\cite{wang2019nsf} and LPCNet~\cite{valin2019lpcnet}.}

The sampling rate of the waveform plays a key role in waveform generation.
It represents the resolution in the time domain when waveforms are being handled on a computer.
The higher the sampling rate, the more accurately the original waveform can be expressed, but the larger will be the amount of data.
With regard to speech, its content and individuality are concentrated in the low frequency range.
Therefore, sampling rates of 16, 22.05, and 24 kHz are often used in research on NVs.
In an environment where only synthetic speech is heard, it does not matter even if the human audible range (20--20,000 Hz) is not entirely covered.
However, in situations where synthetic speech is used for conversing with a human, or where it is played simultaneously with the sound of musical instruments in music, the synthetic speech may sound muffled compared to other sounds.
Some studies (Full-band LPCNet~\cite{matsubara2021full} and PeriodNet~\cite{hono2021periodnet}) have tackled this problem by generating 48 kHz waveforms.
However, in general, the higher the sampling rate, the more difficult it is to generate high-quality waveforms because of the need to model long-term dependencies.
Additionally, because conventional NVs target a specific sampling rate, speech with sampling rates lower than that of the target cannot be used for training.
This is because even if we upsample a waveform with such low sampling rates, high-frequency components cannot be recovered.
Although large speech corpora such as the LJSpeech dataset
\footnote{\cameraready{\url{https://keithito.com/LJ-Speech-Dataset/}}}
and LibriTTS~\cite{zen2019libritts} have been recently released, 44.1 or 48 kHz datasets are still limited.

For these reasons, a method that: 1) can properly model speech with high sampling rates, and 2) also use speech with a sampling rate lower than that of the target for training, is required.
In this study, we propose multiple sampling rate (MSR) -NV as a method with the aforementioned characteristics.
MSR-NV can generate speech waveforms step-by-step, starting from a low sampling rate.
More specifically, after generating a waveform with a certain sampling rate, sinc interpolation is performed to upsample the waveform, and a neural network is used to predict the residual high-frequency components to generate a waveform with a higher sampling rate.
This would allow different networks to capture the features contained in each frequency band and efficiently model waveforms with high sampling rates.
Additionally, MSR-NV enables us to train a part of the model using speech with a low sampling rate.
As a result, it is expected that speech with sampling rates that could not be used in the past can be used together for training, which leads to the realization of a more general-purpose NV.

We conducted experimental evaluations using a model structure based on Parallel WaveGAN (PWG)~\cite{yamamoto2020parallel} to demonstrate the effectiveness of the proposed method.
First, we compared the subjective quality of the speech generated at 16, 24, and 48 kHz using the proposed method with that of the speech generated using the baseline model that was trained separately at these sampling rates.
To investigate the data efficiency of the proposed method, we evaluated the quality of the synthetic speech when the amount of training data was varied from 1 min to 8 h.
Finally, we confirmed that the quality of synthetic speech can be improved using speech with lower sampling rates in conjunction with the original training data.

\mysection{Related works}
\label{sec:related}

Obtaining a high-resolution output is a common challenge not only for speech waveform generation, but also for image generation.
Progressive growing GAN~\cite{karras2018progressive} enables the generation of images with unprecedented 1024$\times$1024 pixels by the gradual addition of layers corresponding to higher resolutions.
The later published StyleGAN~\cite{karras2019style} enables the generation of high-resolution images without changing the topology of the network by preparing layers for multiple resolutions in advance.
Each layer of the network can represent features corresponding to different resolutions using these methods, which can handle multiple resolutions in a stepwise manner.

Several methods have been proposed for speech waveform generation to handle multiple sampling rates.
MelGAN~\cite{kumar2019melgan} and HiFi-GAN~\cite{kong2020hifigan} are methods that can predict waveform by repeatedly upsampling and transforming features. 
HiFi-GAN achieves a quality comparable to natural speech at 22.05 kHz.
VocGAN~\cite{yang2020vocgan} has a generator similar to that of MelGAN, but generates and evaluates waveforms with $\times 1/n~(n=2, 4, 8, 16)$ sampling rates.
However, while the proposed method directly upsamples the waveform and predicts the residual high-frequency components, VocGAN upsamples the features; thus, when a waveform is viewed at multiple sampling rates, the residual structure is not explicitly used.
Additionally, when a high sampling rate (e.g., 44.1, 48 kHz) that covers the entire human audible range is targeted, it is difficult to use speech data with lower sampling rates (e.g., 16, 22.05 kHz) for training.
Even if we upsample them, high-frequency components cannot be recovered.

\mysection{MSR-NV}
\label{sec:prop}

\subsection{Sequential waveform generation of multiple sampling rates}
\label{sec:prop_outline}

\begin{figure}[t]
\centering
\includegraphics[width=\linewidth]{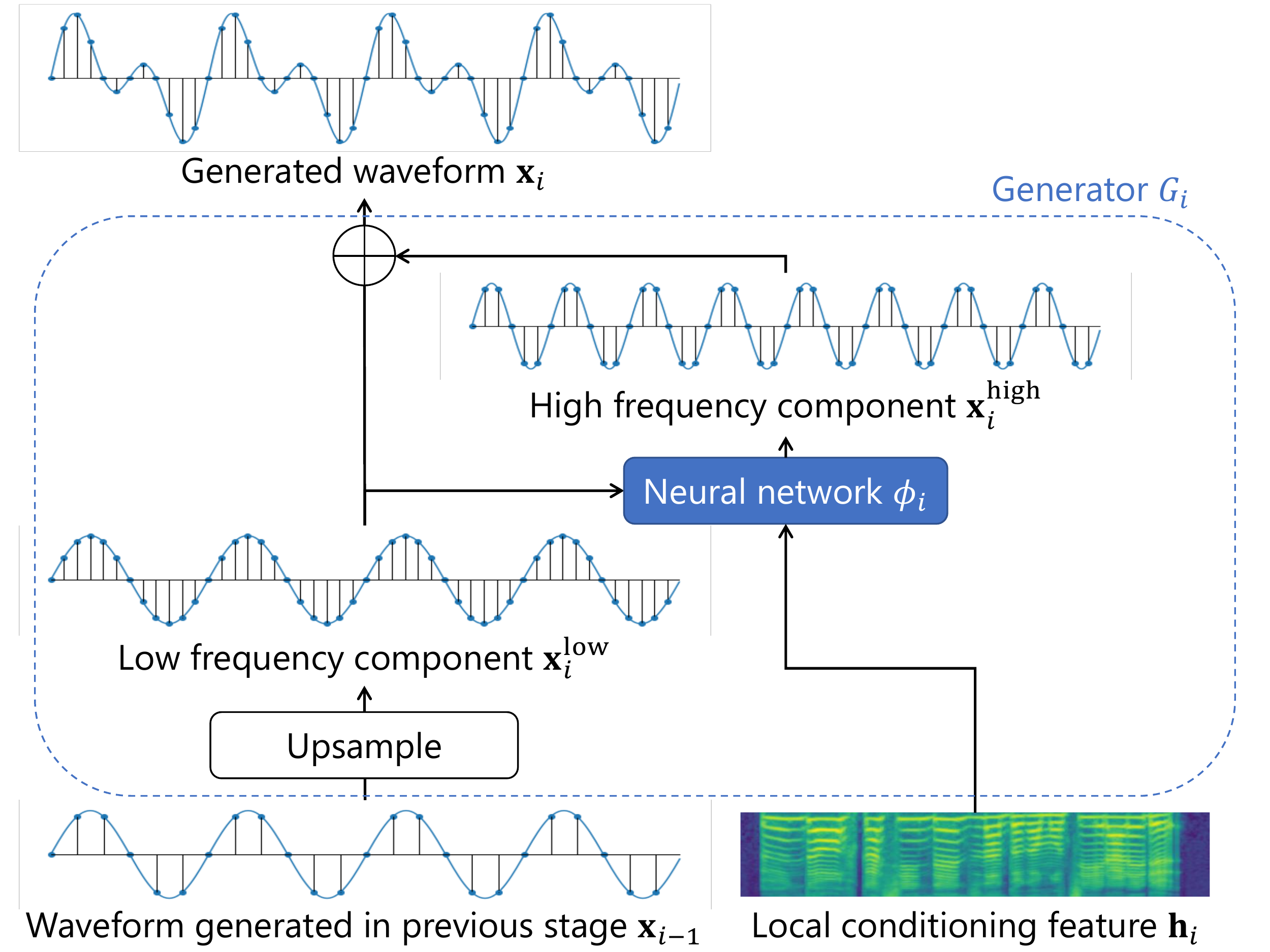}
\caption{Algorithm of generating waveforms of multiple sampling rates step-by-step.}
\label{fig:algorithm}
\vspace{-10pt}
\end{figure}

In this section, we describe the procedure for generating speech waveforms at multiple sampling rates using the proposed method.
Let $f_1<\dots<f_i<\dots<f_I$ be the sequence of the sampling rates to be generated.

First, we consider $i\geq2$, where we predict the waveform $\xB_i$ with a sampling rate $f_i$ from waveform $\xB_{i-1}$ with a sampling rate $f_{i-1}$.
From the linearity of the Fourier transform, $\xB_i$ can be expressed as the sum of a waveform $\xB_i^{\mathrm{low}}$ with a component at frequency $[0, f_{i-1}/2]$ and a waveform $\xB_i^{\mathrm{high}}$ with a component at frequency $[f_{i-1}/2, f_i/2]$.
Therefore, by repeating the following procedure, we can sequentially generate a waveform with a higher sampling rate starting from a low sampling rate:

\begin{enumerate}
\item Upsample $\xB_{i-1}$ to sampling rate $f_i$ to approximate $\xB_i^{\mathrm{low}}$
\item Based on the upsampled waveform $\xB_i^{\mathrm{low}}$ and the conditioning feature $\hB_i$ \cameraready{(\textit{e.g.} upsampled melspectrogram)}, predict $\xB_i^{\mathrm{high}}=\phi_i(\xB_i^{\mathrm{low}}, \hB_i)$ using a neural network $\phi_i$
\item Calculate the sum of 1 and 2: $\xB_i = \xB_i^{\mathrm{low}} + \xB_i^{\mathrm{high}}$
\end{enumerate}
The module that performs the abovementioned procedure is referred to as Generator $G_i$.
\Fig{algorithm} illustrates these steps.
Regarding step~1, transposed convolution is often used for upsampling in NVs~\cite{kumar2019melgan, yang2020vocgan, kong2020hifigan}.
However, this requires that $f_i$ is divisible by $f_{i-1}$ and may include high-frequency components in the range $[f_{i-1}/2, f_i/2]$ that were not included before upsampling.
For these reasons, the proposed method uses upsampling based on sinc interpolation.

Then, we consider the case where $i=1$.
When generating the waveform $\xB_1$ of the lowest sampling rate $f_1$, the previous waveform $\xB_0$ does not exist.
Thus, the function $\phi_1$ predicts $\xB_1=\phi_1(\hB_1)$ with only the conditioning feature $\hB_1$ as the input.
In the proposed method, $f_1$ can be set arbitrarily.
According to Oura et al.~\cite{oura2019deep} and Hono et al.~\cite{hono2021periodnet}, a sinusoidal input corresponding to the fundamental frequency of speech is effective in predicting the periodic component of speech.
Therefore, we set $f_1$ such that the fundamental frequency is included in $[0, f_1/2]$ to make $\xB_1$ similar to the sine wave corresponding to the fundamental frequency.
We expect that $\xB_1$ will facilitate waveform generation in the subsequent stages.

\subsection{Model details}
\label{sec:prop_architecture}

\begin{figure}[t]
\centering
\includegraphics[width=\linewidth]{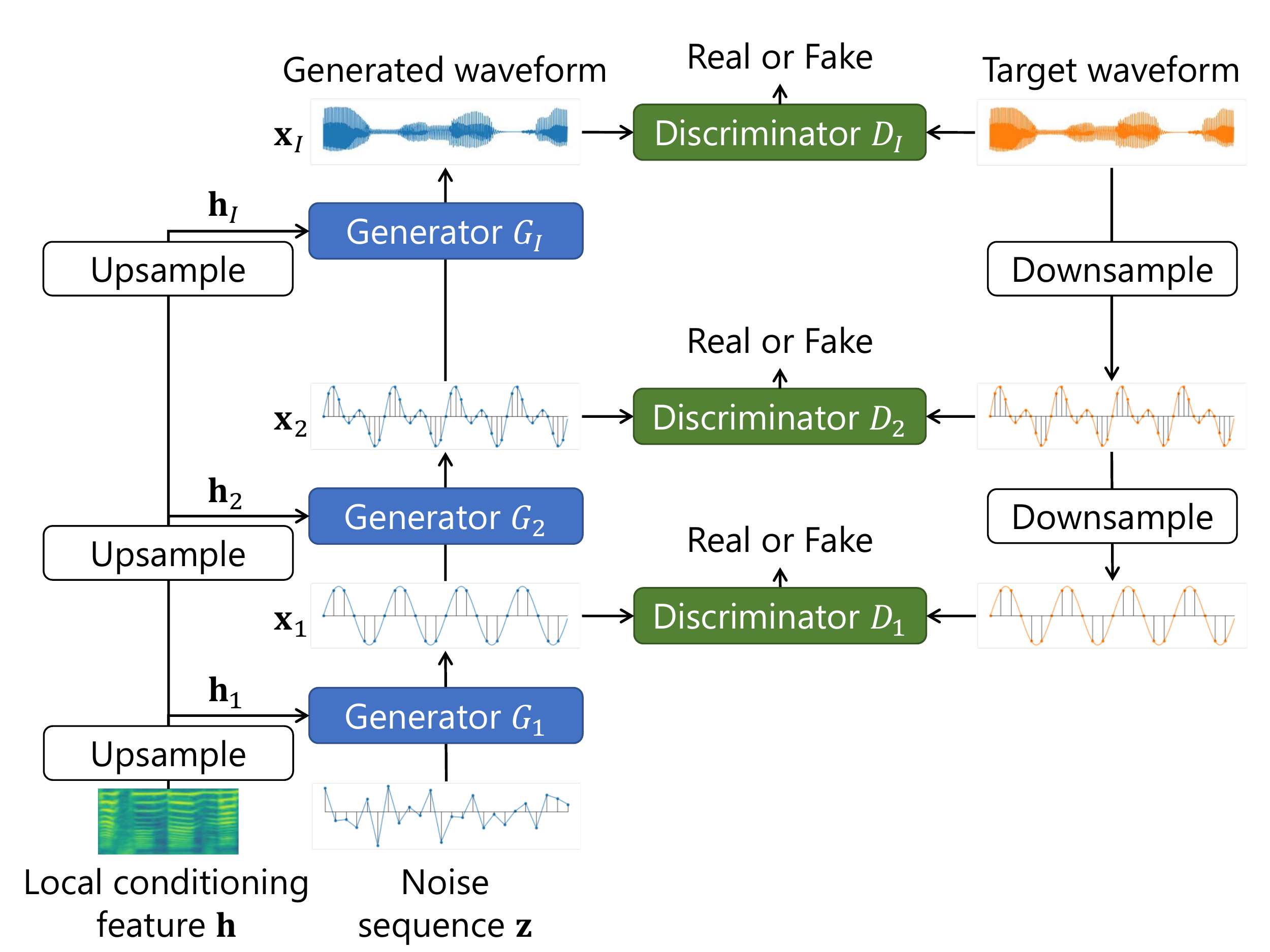}
\caption{Parallel WaveGAN-based model structure extended using the proposed method.}
\label{fig:architecture}
\vspace{-10pt}
\end{figure}

The proposed method described in \Sec{prop_outline} can be applied to NVs of various structures and enables training at high sampling rates, which is difficult with conventional NVs.
In the case of $I=1$, the model structure matches that of a conventional NV, and in the case of $I>1$, it can be regarded as an extension using the proposed method.
We used a model structure based on PWG~\cite{yamamoto2020parallel}.
Although PWG has undergone several improvements since it was first introduced~\cite{wu2021qppwg, song2021improved, yamamoto2021parallel, mizuta2021harmonic, hwang2021high}, we used the original PWG to verify the effectiveness of the proposed method with the simplest model structure.
A conceptual model is depicted in \Fig{architecture}.
As the function $\phi_i$ described in \Sec{prop_outline}, we use the same structure as the generator of PWG, that is, a non-causal WaveNet.
To predict $\xB_i^{\mathrm{high}} = \phi_i(\xB_i^{\mathrm{low}}, \hB_i)$ with a WaveNet-based structure, we must match the temporal resolution of $\xB_i^{\mathrm{low}}$ and $\hB_i$.
Thus, we obtain $\hB_i$ by properly upsampling the original conditioning feature $\hB$ with sinc interpolation.
In the case of $i=1$, we follow PWG and predict $\xB_1=\phi_1(\zB, \hB_1)$ using white noise $\zB$ as input, in addition to the conditioning feature $\hB_1$.

The proposed method generates $I$ waveforms from $\xB_1$ to $\xB_I$. 
Therefore, we use $I$ discriminators $\{D_i\}_{i=1}^I$ to identify whether the corresponding waveform is real or fake.
All the discriminators have identical structure to that of PWG.
A multi-scale discriminator (MSD)~\cite{wang2018high} has also been proposed as a structure to identify downsampled waveforms.
However, MSD loses high-frequency components owing to low-pass processing associated with average pooling.
Thus, we downsampled the natural speech in advance using an anti-aliasing filter and sinc interpolation, and used it as a target at each sampling rate.

\cameraready{We simply extend the loss function used in PWG~\cite{yamamoto2020parallel} to multiple sampling rates.}
For each sampling rate $f_i$, we add multi-resolution short time Fourier transform (MR-STFT) loss $\LM_{\mathrm{aux}}(G_i)$ to adversarial loss $\LM_{\mathrm{adv}}(G_i, D_i)$ weighted by $\lambda_{\mathrm{adv}}$, and use the sum of these as the loss function $\LM_G(G, D)$ for the entire generator.
For the discriminator, we use the sum of the losses $\LM_{\cameraready{\mathrm{dis}}}(G_i, D_i)$ for the discrimination results of $D_i$ as the loss function $\LM_D(G, D)$ for the entire discriminator.
The above can be expressed as follows:
\begin{align}
    \LM_{G}(G, D) &= \sum_{i=1}^I \left\{ \LM_{\mathrm{aux}}(G_i) + \lambda_{\mathrm{adv}}\LM_{\mathrm{adv}}(G_i, D_i) \right\} \label{eq:G}\\
    \LM_{D}(G, D) &= \sum_{i=1}^I \LM_{\cameraready{\mathrm{dis}}}(G_i, D_i) \label{eq:D}
\end{align}
When we use speech with a sampling rate $f_J~(J < I)$ lower than $f_I$ for training, we can train a part of the network by changing the range of summation in Eqs. (\ref{eq:G}) and (\ref{eq:D}) from $i=1,\dots,I$ to $i=1,\dots,J$.

\mysection{Experiments}
\label{sec:exp}

\subsection{Experimental conditions}

\revise{
Experiments were conducted in two settings: single-speaker and multi-speaker.
In the single-speaker experiment, a total of 14,375 utterances (approximately 8 h) uttered in a normal speaking style by a female Japanese speaker were used.
For the development and evaluation sets, respectively, 300 utterances were randomly selected, and the remaining 13,775 utterances were used as the training set.
In the multi-speaker experiment, a total of 31,936 utterances (approximately 25 h) by 26 Japanese speakers (18 female and 8 male), including three speaking styles (normal, happy, and sad) were used.
A total of 520 utterances, 20 from each speaker, were used for the development and evaluation sets, respectively, and the remaining 30,896 utterances were used for the training set.
}
All the experiments were conducted using 48~kHz/16~bit speech signals.
They were trimmed so that the silent interval before and after the speech was approximately 200 ms.
Eighty-dimensional log mel spectrograms with bands in the range 80--7,600 Hz were used as the conditioning feature $\hB$ described in \Sec{prop_architecture}.
They were extracted with a frame and window length of 2,048 points (approximately 42.7 ms) and a frame shift of 240 points (5 ms) and were normalized to zero mean and unit variance.

The sampling rates handled using the proposed method were set to $I=7$ and $\{f_i\}_{i=1}^{7} = \{1, 2, 4, 8, 16, 24, 48\}$ kHz.
$f_1$ was set to 1 kHz so that the fundamental frequency would be included in $[0, f_1/2]$ in most of the frames.
Hereafter, we denote the experimental conditions for generating 48 kHz speech using the proposed method as \method{MSR-PWG-48k}.
All $\{\phi_i\}_{i=1}^{7}$ have identical structures of a ten-layer, one-stack non-causal WaveNet with dilation set to 1, 2, 4, \dots, 512.
Following the experimental conditions of PWG, the number of channels for the \cameraready{residual} block and skip connection was set to 64, and the size of the convolutional filter was set to three.
Upsampling of waveforms and conditioning features was conducted in a differentiable manner using \revise{torchaudio~\cite{yang2021torchaudio}.}
All discriminators were constructed with the same structure as that of the PWG, that is, ten-layer non-causal dilated convolutions with a leaky ReLU activation function ($\alpha=0.2$).

The parameters of the MR-STFT loss were set to different values for each sampling rate.
For $f_7=48$ kHz, we set the frame length to $\{2048, 4096, 1024\}$ points, the window length to $\{1200, 2400, 480\}$ points, and the frame shift to $\{240, 480, 100\}$ points.
For $\{f_i\}_{i=1}^6$, these parameters were multiplied by $f_i/f_7$.
The weight of adversarial loss $\lambda_{\mathrm{adv}}$ was set to 1.0.
A mini-batch was constructed by randomly clipping speech (0.5 s) from eight utterances for each training step.
Training was conducted for 400,000 steps using RAdam optimizer ($\epsilon=10^{-6}$).
Only the generators were optimized in the first 200,000 steps, and the generators and discriminators were jointly optimized in the following 200,000 steps.
The initial learning rates were set to $10^{-3}$ for both the generator and discriminator, and they were reduced by a factor of 0.5, after 300,000 steps of training.

As baseline models, we trained the original PWG with 16, 24, and 48 kHz speech waveforms, respectively (hereinafter referred to as \method{PWG-\{16,24,48\}k})\footnote{\url{https://github.com/kan-bayashi/ParallelWaveGAN} was used.}.
To extract log mel spectrograms, the frame and window lengths were set to 512, 1,024, and 2,048 points for \method{PWG-16k}, \method{PWG-24k}, and \method{PWG-48k}, respectively.
The frame shift was set to 5 ms for all settings.
For \method{PWG-16k} and \method{PWG-24k}, the parameters of the MR-STFT loss are the same as in the previous study~\cite{yamamoto2020parallel}, and they were doubled for \method{PWG-48k}.

\mysubsection{Evaluation}

\mysubsubsection{Comparison to baseline in terms of quality and speed}

\label{sec:exp1}
\begin{table}[t]
\caption{Results of MOS evaluation on closeness to recorded speech quality with 95\% confidence intervals.}
\vspace{-15pt}
\label{tbl:exp1}
\begin{center}
\begin{tabular}{lccc}\toprule
Dataset & \method{Ref} & \method{PWG} & \method{MSR-PWG}\\\midrule
single/16k & 2.48$\pm$0.16 & 1.50$\pm$0.11 & 2.26$\pm$0.14\\
single/24k & 3.97$\pm$0.15 & 2.21$\pm$0.15 & 3.47$\pm$0.15\\
single/48k & 4.43$\pm$0.14 & 1.95$\pm$0.15 & 3.80$\pm$0.16\\\midrule
multi/16k & 3.11$\pm$0.14 & 2.41$\pm$0.13 & 2.35$\pm$0.12\\
multi/24k & 4.25$\pm$0.13 & 2.50$\pm$0.15 & 3.06$\pm$0.15\\
multi/48k & 4.55$\pm$0.10 & 1.84$\pm$0.12 & 3.29$\pm$0.15\\\bottomrule
\end{tabular}
\vspace{-20pt}
\end{center}
\end{table}

A mean opinion score (MOS) test was conducted to evaluate the subjective quality of synthetic speech.
\revise{
In both of the single-speaker and multi-speaker experiments, each rater evaluated ten sets of 9 speech samples, for a total of 90 samples, where each set consisted of a combination of three sampling rates (16~kHz, 24~kHz, and 48~kHz) and three
methods (reference: \method{Ref}, conventional method: \method{PWG}, and proposed method: \method{MSR-PWG})\footnote{Speech samples are available at the following URL: \url{https://rinnakk.github.io/research/publications/MSR-NV}.}
Speech samples were evaluated in terms of closeness to the quality of the 48~kHz reference samples using a five-point scale from 1 (very far) to 5 (very close).
The raters should be able to distinguish between 16, 24, and 48 kHz speech samples;
however, depending on the playback equipment and the subject's hearing, it may not be possible to distinguish between these different sampling rates.
Thus, we conducted an internal preliminary test and asked twenty raters who were able to distinguish between them to participate in the evaluation.
}

The results are presented in \Table{exp1}.
\revise{Note that the scores of \method{PWG} are lower than those of the original paper~\cite{yamamoto2020parallel} owing to the differences in the experimental condition and evaluation criterion.
In the single-speaker experiment, \method{MSR-PWG} achieved higher scores than \method{PWG} at any sampling rate.}
The score of $\method{PWG-48k}$ was lower than that of $\method{PWG-24k}$, even though it was closer to the reference speech in terms of the sampling rate.
This is because \method{PWG} had difficulty generating waveforms with a high sampling rate, resulting in quality degradation.
However, \method{MSR-PWG} achieved a high score even at 48 kHz, which verifies the effectiveness of stepwise waveform generation starting from a low sampling rate.
\revise{
In the multi-speaker setting, 
the scores of \method{MSR-PWG} were comparable to \method{PWG} at 16~kHz and significantly higher at 24~kHz and 48~kHz.
However, there was a gap between the scores of \method{Ref} and those of \method{MSR-PWG}.
In particular, the speech quality was degraded for male speakers, whose amount of data was relatively small in this experiment.
Improving the performance of multi-speaker multi-style modeling is a future challenge for the proposed method.
}

\revise{
In terms of model size, the numbers of parameters were 1.44M for \method{PWG} and 3.06M for \method{MSR-PWG}. When we expanded \method{PWG} to 70 layers to align the number of its parameters with \method{MSR-PWG}, the training and inference time increased significantly without any quality improvement. Thus, we conclude that not the increased number of parameters but the proposed structure improved the speech quality.
}

\begin{table}[t]
\caption{Real-time factor of inference averaged over 300 utterances.}
\vspace{-15pt}
\label{tbl:inference_speed}
\begin{center}
\begin{tabular}{lccc}\toprule
Method & 16~kHz & 24~kHz & 48~kHz\\\midrule
\method{PWG} & 0.027 & 0.039 & 0.074\\
\method{MSR-PWG} & 0.027 & 0.042 & 0.068\\\bottomrule
\end{tabular}
\vspace{-20pt}
\end{center}
\end{table}

We also measured the training and inference speed of the baseline and proposed methods using NVIDIA Tesla P40.
The training required 103 h for \textbf{PWG-16k}, 153 h for \textbf{PWG-24k}, 252 h for \textbf{PWG-48k}, and 132 h for \textbf{MSR-PWG}.
Although the proposed method handled up to 48 kHz waveform generation, it achieved high-quality synthesis with a training time shorter than that of \textbf{PWG-24k}.
For inference, the real-time factor (RTF), which is the time required to generate a waveform in one second, was measured, and the results obtained are presented in \Table{inference_speed}.
Although the generator of the proposed method consists of 70 layers, compared to 30 in the baseline, the inference time did not show an increase.
This is because the input length is proportional to $f_i$, which is especially short in the early stages of the model, and the computational cost is reduced.

\mysubsubsection{Training data amount and synthesis quality}
\label{sec:exp2}

\begin{table}[t]
\caption{Comparison of MOS for various training data amount with 95\% confidence intervals.}
\vspace{-15pt}
\label{tbl:exp2}
\begin{center}
\begin{tabular}{lc}\toprule
Training data amount & MOS\\\midrule
1 min & 1.49$\pm$0.13\\
3 min & 3.07$\pm$0.15\\
5 min & 3.43$\pm$0.16\\
10 min & 3.50$\pm$0.15\\
30 min & 3.55$\pm$0.16\\
8 h (full data) & 3.75$\pm$0.16\\\bottomrule
\end{tabular}
\vspace{-15pt}
\end{center}
\end{table}

Six models were trained with \revise{single-speaker} data of 1, 3, 5, 10, 30 min, and 8 h (full data).
The quality of the 48 kHz synthetic speech was evaluated using the MOS test in the same manner as in \Sec{exp1}.
The results are presented in \Table{exp2}.
Although the score was significantly low when the amount of training data was only 1 min, it greatly improved when 5 min of training data was used.
This result indicates that the proposed method can generate speech of adequate quality even with 5 min of training data.

\subsubsection{Use of speech data with low sampling rates}
\label{sec:exp3}

\begin{table}[t]
\caption{Comparison of MOS for different training sets with 95\% confidence intervals.}
\vspace{-15pt}
\label{tbl:exp3}
\begin{center}
\begin{tabular}{lc}\toprule
Training set & MOS\\\midrule
\method{1min} & 1.76$\pm$0.14\\
\method{3$\times$1min} & 2.67$\pm$0.15\\
\method{3min} & 3.32$\pm$0.17\\\bottomrule
\end{tabular}
\vspace{-20pt}
\end{center}
\end{table}

We investigated whether the quality could be improved using 16 and 24 kHz speech for training when there is little 48 kHz speech available.
The following three conditions were evaluated \revise{in the single-speaker setting} using the MOS test similarly to \Sec{exp1}:
\textbf{1min}: 1 min of 48 kHz speech was used for training,
\textbf{3$\times$1min}: 1 min of different 16, 24, and 48 kHz speech, for 3 min, was used for training, and
\textbf{3min}: 3 min of 48 kHz speech was used for training.
The results are presented in \Table{exp3}.
\textbf{3$\times$1min} showed a significantly higher score than \textbf{1min}, confirming that the quality of synthetic speech can be improved using speech with a lower sampling rate for training.
\cameraready{Although we used only one minute each of 16 and 24 kHz speech here, the synthesis quality of 48 kHz speech could be further improved by increasing the amount of these lower sampling rate speech data.}
However, because 16 and 24 kHz speech do not contain any components above 8 and 12 kHz, respectively, only a part of the network can be trained with these data.
\cameraready{This is why} the score of \textbf{3$\times$1 min} was lower than that of \textbf{3 min}, which was trained using the same amount of 48 kHz speech.

\section{Conclusions}
\label{sec:conlcusion}

In this study, we proposed MSR-NV, a method to handle multiple sampling rates using a single neural vocoder.
Experimental evaluations using a PWG-based structure demonstrated that the proposed method could generate high-quality waveforms at multiple sampling rates, including 48 kHz, while maintaining a fast generation speed.
\revise{We also demonstrated the data efficiency of the proposed method by varying the training data amount and evaluating the speech quality.
In addition, we could improve speech quality using 16 and 24 kHz speech when the amount of 48 kHz speech was limited.}
The proposed MSR-NV eliminates the need to re-train the model for different sampling rates, which broadens the range of applications.

Future work includes verifying the effectiveness of the proposed method when model structures other than PWG are used.
\revise{It is also necessary to improve the speech quality for multiple speakers and speaking styles to realize a more versatile vocoder.}

\vfill\pagebreak

\bibliographystyle{IEEEtran}
\bibliography{ref/speech, ref/image, ref/ml}

\newpage
\appendix

\onecolumn

\begin{figure}[t]
\centering
\includegraphics[width=1.0\linewidth]{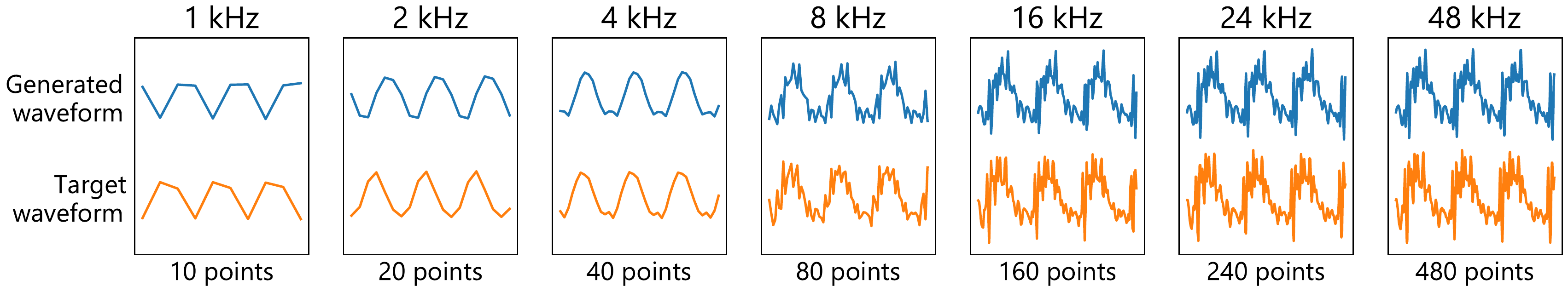}
\caption{Generated waveforms (upper) and target waveforms (lower) of multiple sampling rates.}
\label{fig:waves}
\end{figure}

\begin{multicols*}{2}
\section{Appendix}
\subsection{Waveforms of different sampling rates}
While speech waveforms above 8~kHz are often used in practical applications, speech waveforms below 4 kHz are rarely seen.
The target speech waveforms from 1~kHz to 48~kHz and ones synthesized using the proposed method are shown in \Fig{waves}.
Each waveform corresponds to 10~ms.
We found that the 1~kHz waveform consists of a sinusoidal component corresponding to the fundamental frequency as expected.
We also found that the proposed method could reproduce the shape of the target waveform well at all sampling rates.
On the other hand, phase shifts can be observed because the proposed method does not use the phase information of the target waveform.
This can be avoided by feeding a sine wave obtained from the target waveform~\cite{oura2019deep, hono2021periodnet}; however, we did not adopt that approach because such a system requires fundamental frequency during inference.
Such an approach is considered effective when the proposed method is used for singing voice synthesis.
\end{multicols*}

\end{document}